\documentclass[aps,amsmath,amssymb,floatfix,twocolumn,showpacs]{revtex4}
\usepackage[english]{babel}
\usepackage{amsmath}
\usepackage{amssymb} 
\usepackage{times}
\usepackage{graphicx}
\usepackage{color}

\newcommand{\ket}[1]{|#1\rangle}
\newcommand{\bra}[1]{\langle#1|}

\newcommand{\sbb}{\underline{\sigma}}

\begin{document} 

\title{Exciton Recombination in One-Dimensional Organic Mott Insulators}

\author{Zala Lenar\v ci\v c$^{1}$, Martin Eckstein$^2$, and Peter Prelov\v sek$^{1,3}$}
\affiliation{$^1$J.\ Stefan Institute, SI-1000 Ljubljana, Slovenia}
\affiliation{$^2$Max Planck Research Department for Structural Dynamics, University of Hamburg-CFEL, 22761
Hamburg, Germany}
\affiliation{$^3$Faculty of Mathematics and Physics, University of
Ljubljana, SI-1000 Ljubljana, Slovenia}

\begin{abstract}
We present a theory for the recombination of (charged) holons and doublons in one-dimensional organic Mott insulators, which is responsible for the decay of a photoexcited metallic state. Due to the charge-spin separation, the dominant mechanism for recombination at low density of charges involves a multi-phonon emission. We show that a reasonable coupling to phonons is sufficient to explain the fast recombination observed by pump-probe experiments in ET-F$_2$TCNQ, whereby we can also account for the measured pressure dependence of the recombination rate. 
\end{abstract}


\maketitle

{\bf Introduction --} Femtosecond pump-probe spectroscopy is a powerful probe for the charge relaxation and thermalization phenomena in complex materials. These measurements can directly address and unveil the role of strong electron correlations, as well as the coupling to phonon degrees of freedom. Materials that behave as Mott insulators due to strong electron Coulomb repulsion contain all the latter physics, and are therefore of high theoretical and experimental interest.
It has been observed  that photoinduced charges decay within the picosecond range, i.e. well within the experimental resolution, but on the other hand orders of magnitude faster than in clean semiconductors with similar energy gaps. 

So far two classes of Mott insulators, investigated by pump-probe spectroscopy, revealed similar behavior. These are the layered undoped cuprates La$_2$CuO$_4$ and Nd$_2$CuO$_4$ \cite{matsuda94,okamoto10,okamoto11}, 
and the quasi one-dimensional (1D) organic Mott insulators of the TCNQ family \cite{uemura08}, 
in particular ET-F$_2$TCNQ \cite{okamoto07,uemura08,wall11,mitrano14} which will be the focus of our study. Both undoped cuprates and ET-F$_2$TCNQ reveal ultrafast picosecond charge recombination with some similarities:
(a) The charged carriers created by the pump pulse above the Mott-Hubbard (MH) gap are holons and doublons, and their recombination requires the distribution of a large energy quantum (the MH gap $\Delta \sim 1$eV) into several final excitations with smaller energy $\epsilon_0$.  At low density of charges candidates for recipient bosons can be spin or phonon excitations. 
(b) The decay is exponential in time. This excludes bi- and higher-molecular processes involving inelastic collision of several 'free' charge carriers, and implicitly reveals the existence of an intermediate bound state of a holon and a doublon (the MH exciton). 
In this respect a different observation has been obtained on Ca$_2$CuO$_3$ from the 1D cuprate family, 
which is known to have negligible excitonic effects \cite{ono04} and thus shows a non-exponential decay \cite{matsuzaki15}.

From a theoretical viewpoint the challenge of understanding the charge recombination has analogies with the decay of the double-occupancy in ultracold bosons \cite{chudnovskiy12} and fermions \cite{strohmaier10,sensarma10} in optical lattices, where the decay rate $\Gamma$ exhibits an exponential dependence on the ratio of the Coulomb repulsion $U$ and the typical excitation's energy scale $\epsilon_0$. 
In the latter case the system can be described by a high-temperature state with a sufficient density of excited charges, so that the creation of particle-hole pairs in the compressible background \cite{sensarma10} is the dominant decay channel, and $\epsilon_0$ is set by the kinetic energy of recipient excitations, as observed also within DMFT \cite{eckstein11}.
On the other hand, in real materials the final effective temperature is low, $T\ll U$.
For the case of 2D undoped cuprates, which are antiferromagnets at low $T$, a theory has been presented \cite{lenarcic13,lenarcic14} where the fast charge recombination is explained via emission of spin excitations with the spin exchange energy, $\epsilon_0 \sim J$, as the relevant excitation scale. Strong correlations and large $J$ at the same time lead to a nontrivial origin of the s-type bound state of holon and doublon \cite{tohyama04}, i.e. the MH exciton, being the intermediate state essential for the exponential decay. 

In spite of similarities with 2D Mott insulators, in quasi-1D Mott insulators the scenario involving spin-excitations cannot be effective neither for the MH exciton formation nor for the multi-boson emission due to the phenomenon of charge-spin separation. In the following we will show that a multi-phonon emission can be a viable recombination mechanism in 1D organic Mott insulators, somewhat specific to organic materials with energetic intra-molecular vibrations and strong electron-phonon coupling \cite{girlando11,matsueda12}. The mechanism bears similarity with recently proposed multi-phonon exciton decay in semiconducting carbon nanotubes \cite{perebeinos08} while in standard semiconductors such a scenario seems to be inefficient \cite{yu99}. The prerequisite is again the existence of the 1D MH exciton \cite{essler01}, which can be stable in the case of longer range Coulomb repulsion. Since the photoexcited exciton is of the odd symmetry we will show that its decay becomes allowed only due to the electron-phonon coupling. Finally we show that our scenario can explain the pressure dependence of the recombination rate established recently for ET-F$_2$TCNQ \cite{mitrano14}.

The problem is tackled as follows. By neglecting the recombination term of the Hamiltonian we first compute the exciton, i.e., the lowest bound state in the sector with one doublon and one holon. We then use the Fermi's golden rule in order to compute the decay of the exciton $\ket{\Psi_{1}}$ into the manifold of states $\ket{\Psi_{0}^m}$ that consists of the charge ground state with additional phonon excitations, which in the process of recombination receive the energy of the exciton. 
In principle virtual hoppings give rise to the spin exchange $J = 4t^2/U$, however, we shall neglected it in our calculation. Such approximation is justified by the charge-spin separation specific for 1D, which makes the scattering of charges on spins ineffective
(since the hopping of holons and doublons only shifts the spin background), and by the hierarchy of energy scales $J\ll\omega_0,t,V,U$ for typical organic materials. 

If the Mott gap $\Delta$ is of the order of several phonon frequencies $\omega_0$, $\Delta \approx n\omega_0$, the $n$-phonon contribution determines the matrix element in the Fermi's golden rule expression. The electron phonon-problem is controlled by two dimensionless parameters, the coupling strength $\xi=\lambda^2/\omega_0^2$ and the adiabaticity $t/\omega_0$. 
In the most general case, computing the exciton in the presence of electron-phonon interaction is not possible analytically. To generate admixture of $n\gg1$ phonons to the exciton, the coupling strength  $\xi$ must be treated to higher orders. 
On the other hand, at least the limit $t/\omega_0 \ll 1$ is a valid starting point 
for molecular vibrations in organic crystals. In this case it is convenient to rewrite the Hamiltonian using a unitary Lang-Firsov transformation $e^S$, which measures the phonon coordinate with respect to the equilibrium position for a given charge configuration.
If the transformed exciton state $\ket{\tilde \Psi_1}$ is expanded in phonon number states, $\ket{\tilde \Psi_1} \equiv \ket{\tilde \Psi_1^{(0)}} + \ket{\tilde \Psi_1^{(1)}} + \ket{\tilde \Psi_1^{(2)}} + \ \dots$,  
the zero-phonon state $\ket{\tilde \Psi_1^{(0)}}$ is already the leading contribution in $t/\omega_0$
so that additional phonon-dressing can be neglected. 
Using this approximation we will derive a compact expression for the recombination rate $\Gamma$, 
\begin{align} \label{EqGamma1DSimple-4}
\Gamma
=4\tilde t^2 
&\Big(\frac{1}{2}-\frac{2\tilde{t}^2}{\tilde{V}^2}\Big)  
\sqrt{\frac{2\pi}{\Delta\omega_0}}
\ \times \notag \\
&\times\exp\Big(-\frac{\Delta}{\omega_0}\ln\big(\frac{\Delta}{2e\xi\omega_0}\big)\Big) 
\Big[
1-
(\tfrac12)^{\frac{\Delta}{\omega_0}}
\Big],
\end{align}
which can easily be compared with the experiments, taking the hopping $\tilde t$ and the nearest neighbor interaction $\tilde V$ from independent measurements.

{\bf The Model --} 
As a model for the charge recombination in organic Mott insulators we consider the 1D extended Hubbard model, where in addition to the local Hubbard repulsion $U$ and the nearest-neighbor electron hopping, 
a nearest-neighbor Coulomb repulsion $V>0$ is included. The latter is essential to stabilize the exciton state in 1D \cite{essler01}. 
The Hamiltonian is split in the hopping $H_t$ of doublons and holons, the recombination term $H_{rc}$ and the  interaction term $H_U$, which are written as
\begin{align}
& H_t=-t \sum_{\langle ij\rangle,s} (d_{i,s}^\dagger d_{j,s} - h_{i,s}^\dagger h_{j,s} + \textrm{H.c.}),  \label{EqHt}\\
& H_{rc}=-t \sum_{\langle ij\rangle,s} (h_{i,\bar{s}} d_{j,s} + h_{j,\bar{s}} d_{i,s} + \textrm{H.c.}), \label{EqHrc}\\
&H_U=U \sum_i \frac{n_{i}^d + n_{i}^h}{2} + V \sum_{\langle ij\rangle } \bar n_i \bar n_j, \label{EqHu}
\end{align}
with holon and doublon creation operators 
$
h_{i,s}^\dagger = c_{i,s} (1-n_{i,\bar{s}}), \  
d_{i,s}^\dagger = c_{i,\bar s}^\dagger n_{i,s}
$,  
holon and  doublon density operators $n_{i}^h=\frac{1}{2}\sum_{s} h_{i,s}^\dagger h_{i,s}, n_{i}^d=\frac{1}{2}\sum_{s} d_{i,s}^\dagger d_{i,s}$, and  $\bar n_i=n_i^d-n_i^h$. Here $\langle ij\rangle$ denotes nearest neighbor pairs and $\bar s$ the spin opposite to $s$. 
In addition a generally nonlocal coupling between the charge density and dispersive phonons is introduced,
\begin{align}
&H_{ep}
=\sum_{j,q} \lambda_qe^{-iqj}(a_q^\dagger+ a_{-q})\bar n_j, \
H_{ph}=\sum_q\omega_q a^\dagger_q a_q. \label{hep}
\end{align}

{\bf Lang-Firsov transformation--}  
The derivation of the standard Lang-Firsov transformation for the present case follows Ref.~\cite{lenarcic14}
and is presented in the Supplemental Material.
The exact transformed Hamiltonian is given by
\begin{align}
\tilde H_0 
&=\,
-\tilde t
\sum_{\langle ij \rangle,s}
\Big(
d_{is}^\dagger d_{js} e^{A_{ji}^\dagger} e^{A_{ij}} 
-
h_{is}^\dagger h_{js} e^{A_{ij}^\dagger} e^{A_{ji}} 
+\textrm{H.c.}
\Big)
\nonumber
\\
&\,\,\,\,\,\,\,
+
\tilde U \sum_{j} \frac{n_j^d+n_j^h}{2}
+
\tilde V \sum_{\langle ij\rangle }  \bar n_i \bar n_j
+\sum_{q} \omega_q a_{q}^\dagger a _q,
\\
\tilde H_{rc} 
&=
-\tilde t
\sum_{\langle ij \rangle,s}
\Big(
h_{is} d_{j\bar s} e^{A_{ji}^\dagger} e^{A_{ij}} 
+
h_{js} d_{i\bar s} e^{A_{ij}^\dagger} e^{A_{ji}} 
+\textrm{H.c.}
\Big),\label{EqHrcl}
\end{align}
where 
$
A_{jj'} ^\dagger=\sum_{q}
(\lambda_{q}/\omega_q) (e^{-iqj'}-e^{-iqj})a_q^\dagger
$
is a phonon creation term, and $\tilde U$, $\tilde V$, and $\tilde t$ 
are renormalized interaction and hopping parameters. Corrections to the bare $U,V$ are given by 
$\tilde{U}=U-\tilde{\epsilon}_{0}$ and $\tilde{V}=V-\tilde{\epsilon}_{1}$ with
$\tilde\epsilon_{i-j}=2 \sum_q (|\lambda_q|^2/\omega_q)\cos(q(i-j))$,
 while longer range interaction shall be neglected.
Below we will express all results in terms of the renormalized parameters $\tilde U$, $\tilde V$, and $\tilde t$, which are determined experimentally by a fit to the linear absorption spectrum.

{\bf Exciton ground state --}
We now construct the ground state for the Hamiltonian $\tilde H_0$. 
To neglect additional phonon dressing
as explained above, $\tilde H_0$ is projected 
to the phonon vacuum, $\tilde H_0^{(0)}\equiv |0_{ph} \rangle \langle  0_{ph} | \tilde H_0 | 0_{ph} \rangle \langle 0_{ph}|$. We first construct a basis of all holon-doublon states with an arbitrary spin configuration of the remaining sites, analogous to the squeezed spin state \cite{ogata90}. For a given spin configuration $\sbb = \{\sigma_1,...,\sigma_{L-2}\}$ (with $\sigma_j=\uparrow,\downarrow$), we define $\ket{\sbb^{m,j}}$ as the state obtained by distributing the spins $\sbb$ on lattice sites $\{1,...,L\}  \setminus \{m,j \}$. For $m<j$
\begin{align}
\ket{\sbb^{m,j}}
=&
c_{1,\sigma_1}^\dagger
\cdots
c_{m-1,\sigma_{m-1}}^\dagger
\,\times \,
c_{m+1,\sigma_{m}}^\dagger
\cdots
c_{j-1,\sigma_{j-2}}^\dagger
\,\times \,
\nonumber
\\
&\,\times \,
c_{j+1,\sigma_{j-1}}^\dagger
\cdots
c_{L,\sigma_{L-2}}^\dagger
|0\rangle,
\label{spinstate}
\end{align}
and for $j<m$ analogous. We then define the holon-doublon state 
$\ket{\sbb^m_j}$ by placing a doublon at site $j$,
\begin{equation}\label{dhstate}
\ket{\sbb^m_j} = 
\left\{ \begin{array}{ll}
c_{j,\uparrow}^\dagger c_{j,\downarrow}^\dagger
\ket{\sbb^{m,j}} 
& \textrm{if $m\neq j$},\\
 0 & \textrm{if $m=j$}.
\end{array} \right.
\end{equation}
These are used to define the state with a holon-doublon pair $\ket{\Phi^m_j}$ and the state with two holons $\ket{\Phi^{m,j}}$ on an arbitrary spin background which is a superposition or mixture of configurations $\sbb$, 
\begin{equation}
\ket{\Phi^m_j}\equiv\sum_{\sbb} \Phi_{\sbb}\ket{\sbb^m_j}, \quad
\ket{\Phi^{m,j}} \equiv \sum_{\sbb} \Phi_{\sbb}\ket{\sbb^{m,j}}.
\end{equation}
One can see that the Hamiltonian $\tilde{H}_0^{(0)}$  does not mix different configurations $\sbb$, because nearest neighbor hopping of a holon or doublon implies a shift of the spin background, which is implicit in the definition \eqref{dhstate} for $\ket{\sbb^i_j} \to \ket{\sbb^{i\pm 1}_{j}}$ or ($\ket{\sbb^i_j} \to \ket{\sbb^i_{j\pm 1}}$. The action of the Hamiltonian is thus obtained by
\begin{align}
(\tilde H_0^{(0)}-\tilde{U} + \delta_{|i-j|,1} \tilde V) \ket{\Phi^i_j}
=
-\tilde t
\sum_{\alpha=\pm 1}
\big(
\ket{\Phi^{i+\alpha}_j}
-
\ket{\Phi^i_{j+\alpha}}
\big)
\end{align}
To determine the ground state  we start from a  partial Fourier transform with respect to the average position,
\begin{align}
&\ket{\psi_q^l}=\frac{1}{\sqrt{L}}\sum_{j}e^{iqj+iq(l/2)} \ket{\Phi^j_{j+l}}.
\end{align}
With this the action of the Hamiltonian becomes
\begin{multline}
(\tilde H_0^{(0)} - \tilde U+\delta_{|l|,1} \tilde V)
\ket{\psi_q^l}
=
-
2i
\tilde t_q
\sum_{\alpha=\pm 1}
 \alpha\ket{\psi_q^{l+\alpha}},
\end{multline}
where $\tilde t_q=\tilde t\sin(q/2)$.
There is a continuum of states in the energy window $E\in [\tilde U-4\tilde t_q,\tilde U+4\tilde t_q]$. 
For $\tilde{H}_0^{(0)}$, parity-even and odd bound states are degenerate. We can restrict the analysis to the odd states, which can be created by the optical dipolar transition, and thus make the ansatz
\begin{align}
&\ket{\tilde \Psi_1^{(0)}} =\sum_{l>0} \beta_l \big(\ket{\psi_{q}^l} - \ket{\psi_{q}^{-l}}\big)\ket{0_{ph}}, 
\quad \beta_l=\beta_0 e^{-\kappa_1 l}. \label{phi01}
\end{align}
The ground state is found for $q=\pi$ with 
$\beta_l=\beta_0 (2 \tilde{t}/\tilde{V})^l$
and 
$E_1 = \tilde U-\tilde V- 4\tilde t^2/\tilde V $, which lies below the continuum 
for $\tilde V>2\tilde t$. Without the electron magnon coupling, the exciton is decoupled from the spin background, i.e., excitons for different spin wave functions are degenerate.  

{\bf Exciton decay --}
Similarly to the problem of exciton decay in 2D \cite{lenarcic13,lenarcic14} we establish the recombination rate using the Fermi's golden rule 
\begin{equation}
\Gamma=2\pi \sum_{m}|\langle \Psi_{0}^m| \tilde H_{rc} |\tilde \Psi_{1}^{(0)}\rangle|^2 \ 
\delta(E_{0}^{m}-  E_{1}), \label{Eq1DGam}
\end{equation}
for transitions from previously determined exciton $\ket{\tilde \Psi_1^{(0)}}$, Eq.~(\ref{phi01}), into the charge ground state with additional phonon excitations $\ket{\Psi_{0}^m}$ via the recombination operator $\tilde H_{rc}$, Eq.~(\ref{EqHrcl}). If written in an integral form \cite{lenarcic14}, Eq.~\eqref{Eq1DGam} becomes
\begin{align}
\Gamma
&
=2Re \ \langle \tilde \Psi_{1}^{(0)} | \tilde H_{rc} P_0 
\hspace{-0.1cm} \int_0^\infty \hspace{-0.2cm} d\tau \hspace{-0.05cm}
\ e^{i ( \Delta-H_{ph})\tau} 
P_0  \tilde H_{rc} |\tilde \Psi_{1}^{(0)} \rangle,
\label{Eq1DGamma0b}
\end{align}
where $\Delta$ is the charge gap,  $P_0$ is the projection to the zero charge sector, and $H_{ph}$ is the only part of $\tilde H_0$ which is active in the zero charge sector.

We first evaluate the application of $\tilde H_{rc}$ on the exciton. Starting from the expression $\ket{\tilde \Psi_{1}^{(0)}}$, Eq.~(\ref{phi01}),
we can restrict the application of $\tilde{H}_{rc}$ to nearest neighbor terms $l=1$,
\begin{align}
P_0
\tilde 
H_{rc}
\ket{\Phi^j_{j+1}}\ket{0_{ph}}
=
-\tilde t \
S_{j} \ket{\Phi^{j,j+1}} 
\ e^{A_{j+1,j}^\dagger} \ket{0_{ph}},
\end{align}
where
$
S_{j} = c_{j\uparrow}^\dagger c_{j+1\downarrow}^\dagger -  c_{j\downarrow}^\dagger c_{j+1\uparrow}^\dagger
$ creates a spin singlet on sites $j,j+1$, previously occupied by a holon-doublon pair. Similar,
$P_0
\tilde 
H_{rc}
\ket{\Phi^{j+1}_j}\ket{0_{ph}}
=
-\tilde t \
S_{j} \ket{\Phi^{j,j+1}} \ e^{A_{j,j+1}^\dagger} \ket{0_{ph}}.
$
In summary,
\begin{align}
P_0
\tilde 
H_{rc}
\ket{\tilde \Psi_{1}^{(0)}}
&=
-\tilde t \
\frac{\beta_1}{\sqrt{L}}
\sum_{j} i (-1)^{j} 
S_{j} \ket{\Phi^{j,j+1}}
\,\,
\times
\nonumber
\\
&\hspace{0.07\textwidth}\times
\,\,(e^{A_{j+1,j}^\dagger} - e^{A_{j,j+1}^\dagger}) \ket{0_{ph}}.
\end{align}
When this is inserted into Eq.~\eqref{Eq1DGamma0b}, recombination rate is expressed as
\begin{align}
\Gamma
=
\tilde t^2\beta_1^2
\sum_{d}  g_{d} \,\Gamma^{ph}_{d}(\Delta),
\label{eq0001}
\end{align}
with a spin structure factor
\begin{align}
\label{gspin}
g_{d}
=
(-1)^d\frac{1}{L}
\sum_{j}
\bra{\Phi^{j,j+1}} S_{j}^\dagger S_{j+d} \ket{\Phi^{j+d,j+d+1}},
\end{align}
and a phonon emission factor
\begin{multline}
\Gamma^{ph}_{j-j'}(\Delta)
=
2\text{Re}\int_0^\infty \hspace{-0.15cm}d\tau
e^{i\Delta \tau}\,\bra{0_{ph}}
(e^{A_{j'+1,j'}}-e^{A_{j',j'+1}})
\,\times\,
\\
\times e^{-i H_{ph}\tau} 
(e^{A_{j+1,j}^\dagger}-e^{A_{j,j+1}^\dagger})
\ket{0_{ph}}.
\label{bosonfactor}
\end{multline}

{\bf Spin structure factor --}
From Eqs.~\eqref{gspin} and \eqref{spinstate} one can see that $g_0=2$ and $g_1=g_{-1}=1$ for an arbitrary spin configuration. For $d\ge 2$, Eq.~\eqref{spinstate} implies that for any spin configuration $\sbb$, $ (-1)^d\bra{\sbb^{j,j+1}} S_{j}^\dagger S_{j+d} \ket{\sbb^{j+d,j+d+1}}$
equals $1$ if the spins $(\sigma_{j},...,\sigma_{j+d-1})$ form an antiferromagnetic sequence 
$(\uparrow,\downarrow,\uparrow,\downarrow,...)$ or $(\downarrow,\uparrow,\downarrow,\uparrow,...)$,
and $0$ else. We thus have $g_{d\neq 0}=1$ for a perfect N\'eel antiferromagnet, and $g_{d}=0$ for $|d|\ge 2$ for a spin-polarized background. For a general finite temperature state we expect an exponential decay of the correlations with distance.

{\bf Boson emission factor --}
The matrix element in the boson factor \eqref{bosonfactor} can be evaluated straightforwardly, which is done in the Supplemental Material. We obtain
\begin{align}
\Gamma^{ph}_{d}(\Delta)
&=
8\,\text{Re}\!\!\int_0^\infty \!\!\!d\tau\,
e^{i\Delta \tau}
\,\,\times
\label{Eq1DGamma0b-2}
\\
&\times\,\,
\sinh\big(2\sum_{q} \frac{|\lambda_q|^2}{\omega_q^2}\cos(dq)(1-\cos q)e^{-i\omega_q \tau}\big).
\notag
\end{align}
The argument of the $\sinh$ may be written in a convenient way as an integral
$2\int d\omega e^{-i\omega \tau} f_d(\omega)$,
with the boson coupling function
\begin{align}
f_d(\omega)
&=\sum_{q} \frac{|\lambda_q|^2}{\omega_q^2}\cos(dq)(1-\cos q) \delta(\omega-\omega_q).
\label{gddef}
\end{align}
The zeroth and first moments $\eta_d=\int d\omega f_d(\omega)$, $\Omega_d \eta_d=\int d\omega \, \omega f_d(\omega)$ of these functions are related to the phonon-mediated long-range interaction parameters $\tilde{\epsilon}_{d}$ 
via $2\Omega_d  \eta_d = (\tilde{\epsilon}_{d} - \tfrac12 \tilde{\epsilon}_{d-1} - \tfrac12\tilde{\epsilon}_{d+1})$. The time-integration in Eq.~(\ref{Eq1DGamma0b-2}) can be performed numerically for any kind of dispersions $\lambda_q,\omega_q$, however, for a fixed function $f_d$ and $\Delta/\omega_q\to\infty$, one can 
use an argument related to the central limit theorem to 
show that in the lowest order the result depends only on the zeroth moment $\eta_d$, as presented in the Supplemental Material. 
The integral (\ref{Eq1DGamma0b-2}) can then be approximated with 
\begin{align} 
\Gamma_d^{ph}(\Delta)
&=
2\frac{|\eta_d|}{\eta_d}
\sqrt{\frac{2\pi}{\Delta\omega_0}}
\Big(\frac{\Delta}{2e|\eta_d | \omega_0}\Big)^{-\frac{\Delta}{\omega_0}},
\label{EqGamma1DSimple-DD}
\end{align}
where $\omega_0$ is the typical phonon frequency. Expression (\ref{EqGamma1DSimple-DD}) is obtained also by the saddle point approximation for a Gaussian $f_{d}(\omega)$ with the same zeroth moment $\eta_d$ \cite{lenarcic14}.

\begin{figure}[tb]
\begin{center}
\includegraphics[width=0.4\textwidth]{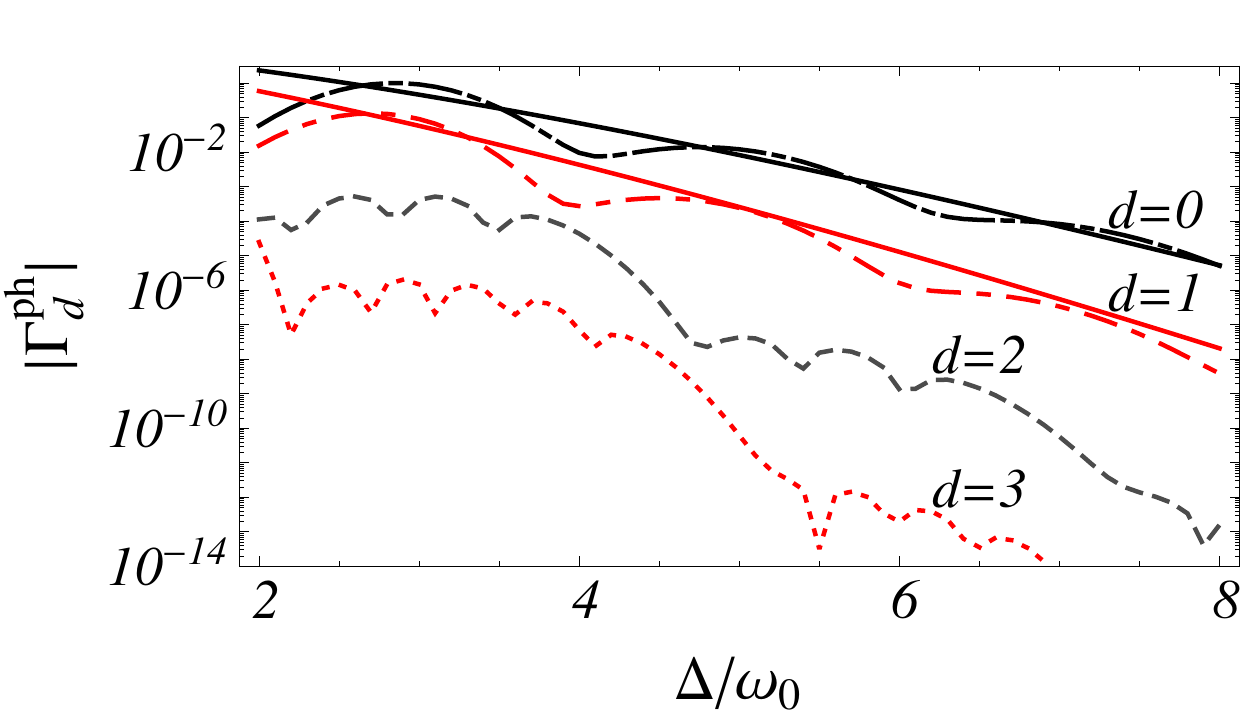}
\end{center}
\caption{(Color online)  Comparison of boson emission factors $\Gamma_{d}^{ph}$ for different $d$, obtained by numerical integration of Eq.~(\ref{Eq1DGamma0b-2}) for dispersions 
$\omega_q=\omega_0 + \delta \omega \cos(q)$, $\lambda_q=\lambda/\sqrt{L}$ (dashed lines) and from Eq.~(\ref{EqGamma1DSimple-DD}) (solid lines). Parameters $\omega_0=1, \delta\omega=0.1 \omega_0, \xi=0.3$ are used.}
\label{Fig1}
\end{figure}

Typically, phonons are weakly dispersive and electron phonon interaction not long-ranged, so that $|\tilde{\epsilon}_{0}| \gg |\tilde{\epsilon}_{1}| \gg |\tilde{\epsilon}_{2}| ...$ Then it suffices to take into account only the $|d|\le 1$ contributions, as demonstrated in Fig.~\ref{Fig1} for dispersions 
$\lambda_q=\lambda/\sqrt{L},$
$\omega_q=\omega_0+\delta\omega\cos(q)$,  
showing results of numerical integration of Eq.~(\ref{Eq1DGamma0b-2}) (dashed lines). To boost the convergence an additional smoothening 
$e^{-i\omega_q \tau} \rightarrow e^{-i\omega_q \tau} e^{-\eta^2 \tau^2/2}$ with $\eta=0.2 \omega_0$ 
has been used in Eq.~(\ref{Eq1DGamma0b-2}), 
which can physically correspond to higher dimensionality of phonons. 
The final expression for the recombination rate, Eq.~(\ref{EqGamma1DSimple-4}), which is relevant for the comparison with experiments, is thus obtained by restricting Eq.~\eqref{eq0001} to the $|d|=0,1$ contributions with spin structure factor $g_0=2, g_{\pm 1}=1$, and using approximations $\Omega_0\approx \Omega_1\approx \omega_0$, $2\omega_0\eta_0 \approx \tilde{\epsilon}_{0}$,  $2\omega_0\eta_1\approx-\tilde{\epsilon}_{0}/2$ in Eq.~(\ref{EqGamma1DSimple-DD}) with $\tilde{\epsilon}_{0}\approx 2\omega_0 \xi$ expressed via coupling strength $\xi=\lambda^2/\omega_0^2$. 
Dependence $\Gamma_d^{ph}(\Delta)$, Eq.~(\ref{EqGamma1DSimple-DD}), with $\eta_d$ approximated as above is for relevant terms $d=0,1$ shown in Fig.~\ref{Fig1} (solid lines), displaying agreement with numerical integration. 
The prefactor $(1/2-2\tilde{t}^2/\tilde{V}^2)$ in Eq.~(\ref{EqGamma1DSimple-4}) comes from $\beta_1^2$.

{\bf Comparison with experiment --}
Finally we compare the recombination times $\tau_r=\Gamma^{-1}$ obtained from Eq.~(\ref{EqGamma1DSimple-4}) with experimentally measured ones \cite{mitrano14}.   
All quantities but the strength of charge-phonon coupling $\xi$ are set by the experimental data: 
$\omega_0=0.23eV$ \cite{kaiser14}, $\tilde U=0.845eV$, while $\tilde{t}(p)\in [0,04,0.06]eV$, $\tilde{V}(p)\in [0.12,0.16]eV$ are specified functions of pressure \cite{mitrano14}, and 
$\Delta\approx \tilde U-\tilde V-4\tilde t^2/\tilde V$. 
\begin{figure}[ht]
\begin{center}
\includegraphics[width=0.4\textwidth]{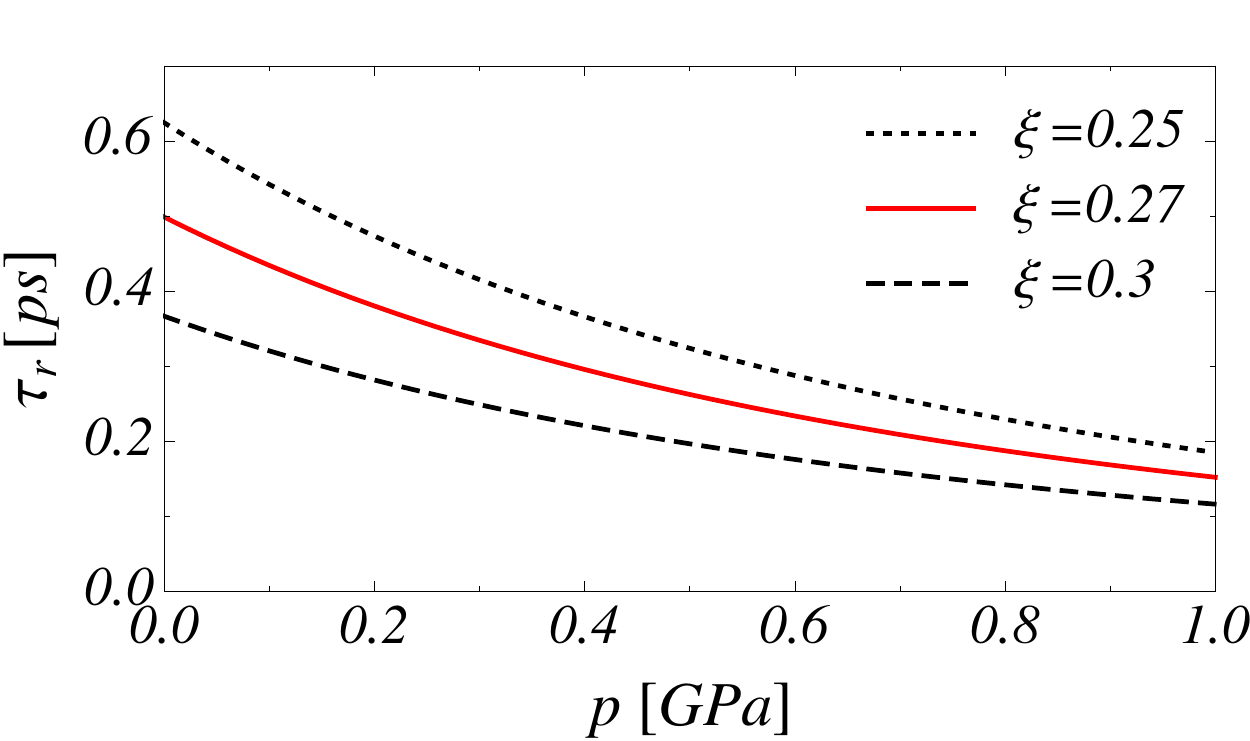}
\end{center}
\caption{(Color online) Recombination time $\tau_r$ as a function of pressure $p$ calculated from the Eq.~(\ref{EqGamma1DSimple-4}) using the experimental parameters \cite{mitrano14,kaiser14}: 
$\omega_0=0.23eV, U=0.845eV$, and $\tilde{t}(p)\in [0,04,0.06]eV$, $\tilde{V}(p)\in [0.12,0.16]eV$ for coupling strengths $\xi=0.25, 0.27,0.3$.}\label{FigExperiment}
\end{figure} 
Fig.~\ref{FigExperiment} displays $\tau_r$ as a function of pressure $p$ for three different values $\xi=0.25,0.27,0.30$, showing that $\xi \approx 0.27$ is consistent with the experimentally measured recombination times \cite{mitrano14}, yielding the electron-phonon coupling
$\lambda=\omega_0\sqrt{\xi}=0.12eV$.
The latter has been measured and calculated for a similar organic material, finding 
$\lambda\in[0.05eV,0.1eV]$ \cite{girlando11}, confirming that the electron-phonon coupling needed to reproduce the experimental results is indeed realistic.

{\bf Conclusions and discussion --}
The central result of our study is that fast charge recombination observed recently in quasi-1D organic Mott insulators \cite{mitrano14} can be explained via creation of phonon excitations, which can be for the material considered (ET-F$_2$TCNQ) identified as molecular vibrations. Due to the charge-spin separation in 1D systems and hierarchy of energies $J<\omega_0$ in materials addressed, spin excitations are in contrast to 2D systems an inefficient decay channel and were neglected in our analysis by setting $J\rightarrow 0$. Motivated by the experimentally observed exponential decay of charge density we derive the recombination rate based on the assumption that holon and doublon initially 
form a bound state - exciton, which is odd under the parity transformation (therefore optically accessible). Still, the transition into the charge g.s. with even parity is allowed due to the coupling to phonons. We established the charge recombination rate using the Fermi's golden rule, showing approximately exponential suppression with the number of phonons emitted in the process; an observation common to several doublon decay processes \cite{strohmaier10,sensarma10,sensarma11,eckstein11,lenarcic13,lenarcic14}  with different recipients of doublon energy.

The experimentally established frequency of the relevant vibrations \cite{kaiser14} is much larger than the one of typical lattice phonons, making the recombination mechanism somewhat specific for organic insulators. To explain recent experiments on 1D cuprates (Ca$_{2}$CuO$_{3}$) \cite{matsuzaki15} with smaller typical phonon frequencies and negligible excitonic effect some modification of the mechanism might be needed and remains as a future challenge. One should note that to assist a proper dissipation of energy in the case considered the vibrations must be at least partially dispersive or coupled to other modes. Even though our derivation focuses on 1D phonons, it is straightforward to generalize it to more realistic three-dimensional electron-phonon coupling. Recognizing the role of the electron-phonon coupling in the recombination mechanism we see the recombination measurements as an indirect way to establish its typically elusive strength at least for this class of materials.

\acknowledgments
The authors acknowledge helpful discussions about the experimental result with S. Kaiser, M. Mitrano, and H. Okamoto. 
This work has been supported by the Program P1-0044 and the project J1-4244 
of the Slovenian  Research Agency (ARRS). Z. L. is supported also be the L'Or\'eal - UNESCO national scholarship "For Women in Science".

\bibliographystyle{physrev4}

\onecolumngrid
\newpage
\begin{center}

{\large\textbf{\boldmath
Supplemental Material\\ [0.5em] to \\ [0.5em]
Exciton Recombination in One-Dimensional Organic Mott Insulators}}\\[1.5em]

Zala Lenar\v ci\v c$^{1}$, Martin Eckstein$^2$ and Peter Prelov\v sek$^3$\\[0.5em]

\textit{\small
$^1$J.\ Stefan Institute, SI-1000 Ljubljana, Slovenia\\
$^2$Max Planck Research Department for Structural Dynamics, University of Hamburg-CFEL, 22761
Hamburg, Germany\\
$^3$Faculty of Mathematics and Physics, University of Ljubljana, SI-1000 Ljubljana, Slovenia
}

\vspace{2em}
\end{center}

\twocolumngrid
\setcounter{figure}{0}
\renewcommand{\thefigure}{SM\arabic{figure}}
\renewcommand*{\citenumfont}[1]{S#1}
\renewcommand*{\bibnumfmt}[1]{[S#1]}
\section{Lang-Firsov Transformation}
\label{LFTRAFO}

The unitary Lang-Firsov transformation $e^{S}$ is obtained by choosing the operator $S$ of the form
\begin{equation}
\label{LFS}
S=\sum_{j,q}\alpha_{q}^j (a_{q}^\dagger - a_{-q}) \bar n_{j},
\end{equation}
i.e., a coherent displacement of the phonon coordinate depending on the holon-doublon configuration. The parameters $\alpha_q^j$ are chosen such that the direct coupling term $H_{ep}$ in the transformed Hamiltonian 
$\tilde H=e^{-S} H e^S$ 
is eliminated, i.e., $H_{ep} - [S,H_{ph}] =0$, which is achieved by  the choice $\alpha_q^j=-\lambda_q e^{-iqj}/\omega_q$. To see this we first write
\begin{align}
S
&=\sum_{q}  (-\frac{\lambda_q  N_q}{\omega_q} \ a_{q}^\dagger + \frac{\lambda_{-q}  N_{-q}}{\omega_q}\ a_{q}),
\end{align}
with 
$N_q = \sum_j e^{-iqj} \bar n_j = N_{-q}^\dagger$. In this representation, the terms for individual momenta commute, so that one can easily compute the transformation of $H_{ph}$ and $H_{ep}$. Using expressions for a  coherent state shift of bosonic operators,
\begin{align}
&
e^{ \alpha a^\dagger  - \alpha^* a }  \,a \,e^{ - \alpha a^\dagger  + \alpha^* a }  = a - \alpha 
\\
&e^{ \alpha a^\dagger  - \alpha^* a }  \,a^\dagger  e^{ - \alpha a^\dagger  + \alpha^* a }  = a^\dagger - \alpha^*,
\end{align}
we have 
\begin{align}
&
e^{-S} a_q e^{ S} =  a_q - \frac{\lambda_q  N_q}{\omega_q}
\\
&
e^{-S} a_q^\dagger  e^{ S} =  a_q^\dagger - \frac{\lambda_{-q}  N_{-q}}{\omega_q}.
\end{align}
Hence, writing Eq.~\eqref{hep} as $H_{ep}=\sum_{q} \big( \lambda_q a_q^\dagger N_{q} + \lambda_{-q} a_q N_{-q}\big)$, we have
\begin{align}
e^{-S}
&
\big(
H_{ph}
+
H_{ep}
\big)
e^{S}
=
\nonumber
\\
&=\sum_{q}
\omega_q 
a_{q}^\dagger
a_q
-
\sum_{q}
\frac{\lambda_q \lambda_{-q}}{\omega_q}
N_{q}N_{-q}
\\
&
=
\sum_{q}
\omega_q 
a_{q}^\dagger
a_q
-
\frac{1}{2}\sum_{jj'}
\bar n_{j} \,\tilde{\epsilon}_{j-j'}\,\bar n_{j'},
\end{align}
where 
\begin{align}
\label{long-range-couplings}
\tilde{\epsilon}_{j-j'} =  2 \sum_q  \frac{|\lambda_q|^2}{\omega_q}
\cos\big(q(j-j')\big)
=
\tilde{\epsilon}_{j'-j}
\end{align}
are phonon-induced long-range interaction parameters. For the transformation of the hopping we introduce a different representation of Eq.~\eqref{LFS},
\begin{align}
&
S = \sum_{j} p_j \bar n_j,
\\
\label{pidef}
&p_j
=
-p_j^\dagger
=
\sum_q
\Big(
-e^{-iqj}
\frac{\lambda_q}{\omega_q} a_{q}^\dagger
+
e^{iqj}
\frac{\lambda_{-q}}{\omega_q} a_{q}
\Big).
\end{align}
It is important to note that the operators $\bar n_j p_j $ and $p_l \bar n_l$ do commute, because 
the density operators commute, and we have
\begin{align}
[p_j,p_{l} ]
&=
2i\sum_{q}
\frac{|\lambda_q|^2}{\omega_q^2} \sin\big(q(l-j)\big),
\end{align}
which vanishes due to $\lambda_q=\lambda_{-q}^*$, $\omega_q=\omega_{-q}$.
With this we have (because  $ \bar n_j d_{js}  = 0 $, $ d_{js}  \bar n_j = d_{js}  $)
\begin{align}
\label{dtransform}
e^{-S}
d_{js}
e^{S}
&=
e^{- \bar n_j p_j}
d_{js}
\,
e^{ \bar n_j p_j}
=
d_{js}
e^{ p_j}.
\end{align}
An analogous calculation for the holon operator gives,
\begin{align}
\label{holetransform}
e^{-S}
h_{js}
e^{S}
=
&
h_{js}
\,
e^{-p_j}.
\end{align}
Equations \eqref{dtransform} and \eqref{holetransform} can now be used to 
transform the hopping part \eqref{EqHt} of the Hamiltonian,
\begin{align}
\label{hhtransform}
&
e^{-S}
h_{js}^\dagger
h_{j's}
e^{S}
=
h_{js}^\dagger
h_{j's}
e^{p_j}
e^{-p_{j'}},
\\
\label{ddtransform}
&
e^{-S}
d_{js}^\dagger
d_{j's}
e^{S}
=
d_{js}^\dagger
d_{j's}
e^{-p_j}
e^{p_{j'}}.
\end{align}
In these expressions, electron-phonon interaction is present through the factors $e^{ p_j}e^{-p_{j'}}$ and $e^{ -p_j}e^{p_{j'}}$. 

Below we will study the action of the Hamiltonian  $\tilde  H_0 $, in particular within the zero-phonon sector. For this purpose, it is convenient to rewrite all phonon-operators, especially the terms $e^{ p_j}e^{-p_{j'}}$ and $e^{ -p_j}e^{p_{j'}}$ in Eqs. \eqref{ddtransform} and \eqref{hhtransform}, in a normal-ordered form, so that they give zero when acting on the phonon vacuum. With Eq.~\eqref{pidef} we define
\begin{align}
p_j
&
\equiv A_{j}^\dagger - A_{j},
\,\,\,
A_j ^\dagger
=
-\sum_{q}
e^{-iqj}
\frac{\lambda_{q}}
{\omega_q} a_q^\dagger.
\label{Ajdef}
\end{align}
Using the Baker Hausdorff relation 
$e^{X+Y} = e^{X}e^Y e^{-\frac{1}{2}[X,Y]}$ and 
$e^{X}e^{Y} = e^{Y}e^X e^{[X,Y]}$ one then gets
\begin{align}
\label{phonhopping}
e^{p_j}
e^{-p_{j'}}
&=
e^{A_{jj'}^\dagger}
e^{-A_{jj'}}
e^{-\xi_{j-j'}},
\\
\label{phonhopping02}
e^{-p_j}
e^{p_{j'}}
&=
e^{-A_{jj'}^\dagger}
e^{A_{jj'}}
e^{-\xi_{j-j'}},
\\
\label{Aijdef}
A_{jj'} &= A_j -A_{j'},
\\
\label{xifac01}
\xi_{j-j'}
&=
[A_{j'}^\dagger,A_j] - \frac{1}{2} [A_j^\dagger,A_j]  - \frac{1}{2} [A_{j'}^\dagger,A_{j'}]
\\
\label{xifac}
&=
\sum_{q}
\frac{|\lambda_q|^2}{\omega_q^2}
\big[
1-\cos(q(j-j'))\big].
\end{align}

When the hopping Hamiltonian is projected to the phonon vacuum, only a renormalization of the hopping by a factor $e^{-\xi_{j-j'}}$ (the Franck-Condon factor) remains. Using Eqs.~\eqref{hhtransform}, \eqref{ddtransform}, \eqref{phonhopping}, \eqref{phonhopping02},
and \eqref{xifac} the transformed hopping Hamiltonian $\tilde H_t = e^{-S}H_t e^{S}$ in the Lang-Firsov representation is obtained, and together with the interaction term $\tilde H_{int}$ reads 
$\tilde H_0=\tilde H_t +\tilde H_{int}$,
\begin{align}
&
\tilde H_t 
=
-\tilde t
\sum_{\langle ij \rangle,s} \hspace{-0.15cm}
\Big(
d_{is}^\dagger d_{js} e^{-A_{ij}^\dagger} e^{A_{ij}} 
-
h_{is}^\dagger h_{js} e^{A_{ij}^\dagger} e^{-A_{ij}} 
+\textrm{H.c.}
\Big)
\\
&
\tilde H_{int} = 
\sum_{q} \omega_q a_{q}^\dagger a _q
+
\frac{\tilde U}{2} \sum_{i} \bar n_i^2
+
\frac{1}{2}\sum_{ij} \tilde V_{i-j} \bar n_i \bar n_j
\end{align}
where $\tilde U=U-\tilde{\epsilon}_{0}$ and $\tilde V_l = \delta_{l,1} V-\tilde{\epsilon}_{l}$
and
\begin{align}
\tilde t = t e^{-\xi_1}
\end{align}
are renormalized interaction parameters.  The parameters  $\tilde t$ and $\tilde V_l$ 
are determined experimentally, by characterization of the exciton in linear absorption. For simplicity, we thus include only the nearest neighbor interaction $\tilde V_1\equiv \tilde V$ in the simulation, other parameters are small and experimentally not known, while the calculation  can be straightforwardly extended to longer range hopping.

Similar to Eqs.~\eqref{hhtransform} and \eqref{ddtransform}, the recombination term $\tilde H_{rc}$ reads
\begin{align}
&
\tilde H_{rc} 
=
-\tilde t
\sum_{\langle ij \rangle,s}
\Big(
h_{is} d_{j\bar s} e^{-A_{ij}^\dagger} e^{A_{ij}} 
+
h_{js} d_{i\bar s} e^{-A_{ji}^\dagger} e^{A_{ji}} 
+\textrm{H.c.}
\Big).
\end{align}

\section{Evaluation of the Boson factor}

First we evaluate the integrand of Eq.~\eqref{bosonfactor}. It can be written as (for $d=j-j'$)
\begin{align}
\bra{0_{ph}}
(e^{A_{1,0}(\tau)}-e^{A_{0,1}(\tau)}) (e^{A_{d+1,d}^\dagger}-e^{A_{d,d+1}^\dagger})
\ket{0_{ph}},
\end{align}
where the time argument of $A_{jj'}$ is due to the evolution with $H_{ph}$, which simply amount to 
replacing $a_q$ by $a_q e^{-i\omega_q \tau}$. Using the Baker-Hausdorff relation 
$e^{X}e^{Y} = e^{Y}e^X e^{[X,Y]}$ we can normal-order the bosonic operators, which gives
\begin{multline}
e^{[A_{1,0}(\tau),A_{d+1,d}^\dagger]}
-
e^{[A_{1,0}(\tau),A_{d,d+1}^\dagger]}
\\
+
e^{[A_{0,1}(\tau),A_{d,d+1}^\dagger]}
-
e^{[A_{0,1}(\tau),A_{d+1,d}^\dagger]}.
\end{multline}
The exponentials can be evaluated using Eqs.~\eqref{Ajdef} and \eqref{Aijdef}, yielding
\begin{align}
4\sinh\big(2\sum_{q} \frac{|\lambda_q|^2}{\omega_q^2}\cos(dq)(1-\cos q)e^{-i\omega_q\tau}\big).
\label{Eq1DGamma0b-1-app}
\end{align}

Now we argue how the time integration in Eq.~(\ref{Eq1DGamma0b-2}) can be approximately evaluated.
While the precise form of the boson coupling function $f_d(\omega)$ is often not known, for optical phonons it is centered around some frequency $\omega_0$, and could be approximated with a Gaussian form
\begin{align}
\mathcal{G}_d(\omega) 
=
\frac{\eta_d}{\sqrt{2\pi} \sigma_d}  \ e^{-(\omega-\omega_0)^2/2 \sigma_d^2},
\label{EqGauss}
\end{align}
for which the integral (\ref{Eq1DGamma0b-2}) can be established using saddle point approximation \cite{lenarcic14-2}, yielding decay rate of form (\ref{EqGamma1DSimple-DD}). For a more general coupling function centered at a typical frequency $\omega_0$ an argument related to the central limit theorem can be used to show that in the limit $\Delta/\omega_0\rightarrow \infty$  the decay rate is up to the leading order determined by its zeroth moment $\eta_d=\int d\omega f_d(\omega)$ as written in Eq.~(\ref{EqGamma1DSimple-DD}).

We define 
\begin{align}
&f_d(\tau) \equiv \int_{-\infty}^\infty d\omega e^{-i\omega \tau} f_{d}(\omega), \\
&\tilde{f}_d(\omega) \equiv \frac{1}{\eta_d} f_d(\omega+\omega_0), \\
&\mathcal{F}_d^{(m)}(\omega) \equiv \tilde{f}_d * \dots * \tilde{f}_d \ |_{\omega},
\end{align}
where  $* \dots *$ is the $m$-fold convolution of $\tilde{f}_d(\omega)$.
The expression for the recombination rate Eq.~(\ref{Eq1DGamma0b-2}) can then be simplified as
\begin{align}
\Gamma_d^{ph}(\Delta)
&=8 Re \int_0^\infty \hspace{-0.25cm}d\tau e^{i \Delta \tau} \sinh\Big(2 \int_{-\infty}^{\infty} \hspace{-0.25cm} d\omega e^{-i\omega \tau} f_d(\omega)\Big) \\
&=4 \sum_{m\textrm{ odd}} \frac{2^m}{m!} \int_{-\infty}^\infty \hspace{-0.1cm}d\tau e^{i \Delta \tau} (f_d(\tau))^m \\
&=8\pi \ \frac{|\eta_d|}{\eta_d}\sum_{m \textrm{ odd}} \frac{(2|\eta_d|)^m}{m!} \ \mathcal{F}_d^{(m)}(\Delta-m\omega_0).
\end{align}
The $m$-fold contribution $\mathcal{F}_d^{(m)}(x)$ of the normalized functions becomes broadly centered around $x=0$
for large $m$.  The dominant contributions to the sum will come from terms around $\bar{n}=\Delta/\omega_0$, therefore we can make further approximations
\begin{align}
\Gamma_d^{ph}(\Delta)
&\approx 8\pi \ \frac{|\eta_d|}{\eta_d} \ \frac{(2|\eta_d|)^{\bar{n}}}{\bar{n}!} \ \sum_{m \textrm{ odd}} \mathcal{F}_d^{(n)}(\Delta-m\omega_0).
\end{align}
where $n=\lfloor \bar n \rfloor$ stands for the integer part of $\bar n$. Using that for large $n$ with broad $\mathcal{F}_d^{(n)}$
\begin{align}
\sum_{m \textrm{ odd}}\mathcal{F}_d^{(n)}(\Delta-m\omega_0) 
&\approx \frac{1}{2\omega_0} \int d\omega \mathcal{F}_d^{(n)}(\Delta-\omega) \\
&\approx \frac{1}{2\omega_0}
\end{align}
and the Stirling approximation 
$n!\approx \sqrt{2\pi n} (n/e)^n$,
we finally obtain the compact expression (\ref{EqGamma1DSimple-DD}),
\begin{align}
\Gamma_d^{ph}(\Delta)
&\approx 8\pi \ \frac{|\eta_d|}{\eta_d} \ \frac{(2|\eta_d|)^{\bar{n}}}{\bar{n}!} \ \frac{1}{2\omega_0} \\
&\approx \frac{8\pi}{2\omega_0} \ \frac{|\eta_d|}{\eta_d} \ 
\sqrt{\frac{\omega_0}{2\pi \Delta}} \Big(\frac{\Delta}{e\omega_0}\Big)^{-\frac{\Delta}{\omega_0}} \ 
(2 |\eta_d|)^{\frac{\Delta}{\omega_0}} \\
&= 2 \ \frac{|\eta_d|}{\eta_d} \sqrt{\frac{2\pi}{\Delta \omega_0}} \Big(\frac{\Delta}{2 |\eta_d | e \omega_0}\Big)^{-\frac{\Delta}{\omega_0}}.
\end{align}

\end{document}